\documentclass[preprintnumbers,article,amsmath,amssymb,floatfix,10pt,prd,onecolumn,
superscriptaddress,nofootinbib]{revtex4}
\usepackage{bm}
\usepackage{amsfonts}
\usepackage{latexsym}
\usepackage[latin1]{inputenc}
\usepackage{graphicx}
\usepackage{amsmath}
\usepackage{palatino}
\usepackage{mathpazo}
\usepackage{textcomp}
\usepackage{float}
\usepackage{booktabs}
\usepackage{dcolumn}
\usepackage{ragged2e}
\usepackage{hyperref}
\hypersetup{colorlinks,citecolor=blue}
\hypersetup{colorlinks=true,linkcolor=red,filecolor=magenta,    urlcolor=blue}
\usepackage{amsmath}
\usepackage{xcolor}
\usepackage{epsfig}
\usepackage{caption}
\usepackage{appendix}
\usepackage{subcaption}
\usepackage{booktabs}

\usepackage{commath}
\newcommand{\ba}{\begin{eqnarray}}
\newcommand{\ea}{\end{eqnarray}}

\allowdisplaybreaks[1]

\begin{document}
\color{black}       
%
\title{Cosmological dynamics of interacting dark energy and dark matter in $f(Q)$ gravity}

\author{Gaurav N. Gadbail}
\email{gauravgadbail6@gmail.com}
\affiliation{Department of Mathematics, Birla Institute of Technology and Science-Pilani,\\ Hyderabad Campus, Hyderabad-500078, India.}

\author{Simran Arora}
\email{dawrasimran27@gmail.com}
\affiliation{Center for Gravitational Physics and Quantum Information, Yukawa Institute for Theoretical Physics, Kyoto University, 606-8502, Kyoto, Japan}

\author{Phongpichit Channuie} 
\email{phongpichit.ch@mail.wu.ac.th}
\affiliation{College of Graduate Studies, Walailak University, Thasala, Nakhon Si Thammarat, 80160, Thailand}
\affiliation{School of Science, Walailak University, Thasala, Nakhon Si Thammarat, 80160, Thailand}

\author{P.K. Sahoo}
\email{pksahoo@hyderabad.bits-pilani.ac.in}
\affiliation{Department of Mathematics, Birla Institute of Technology and Science-Pilani,\\ Hyderabad Campus, Hyderabad-500078, India.}

%

\begin{abstract}
In this work, we explore the behavior of interacting dark energy (DE) and dark matter (DM) within a model of $f(Q)$ gravity, employing a standard framework of dynamical system analysis. We consider the power-law $f(Q)$ model incorporating with two different forms of interacting DE and DM: $3\alpha H\rho_m$ and $\frac{\alpha}{3H}\rho_m \rho_{DE}$. The evolution of \(\Omega_m\), \(\Omega_r\), \(\Omega_{DE}\), \(q\), and \(\omega\) for different values of the model parameter \(n\) and the interaction parameter \(\alpha\) has been examined. Our results show that the universe was dominated by matter in the early stages and will be dominated by DE in later stages. Using the observational data, the fixed points are found to be stable and can be represented the de Sitter and quintessence acceleration solutions. We discover that the dynamical profiles of the universe in $f(Q)$ DE models are influenced by both the interaction term and the relevant model parameters.\\


\end{abstract}

\maketitle

\date{\today}

\section{\NoCaseChange{Introduction}}

The $\Lambda$-cold dark matter ($\Lambda$CDM) model has been extensively validated across a wide range of scales, from cosmic to very small, through various observations, including type Ia supernovae (SNe Ia) \cite{SupernovaSearchTeam:1998fmf,Riess:1998dv,SupernovaCosmologyProject:1998vns}, baryon acoustic oscillations (BAO) \cite{SDSS:2005xqv,SDSS:2009ocz}, and the cosmic microwave background (CMB) \cite{WMAP:2003elm,WMAP:2010qai}. While $\Lambda$CDM has successfully explained numerous cosmological phenomena, it still encounters two significant challenges: (i) why does the cosmological constant observed today differ vastly from its theoretical prediction? and (ii) why are the current densities of dark matter (DM) and dark energy (DE) of the same magnitude? DM plays a crucial role in the formation of cosmic structures, while DE is responsible for driving the current accelerated expansion of the universe. While the nature of DM is partially understood through its indirect gravitational effects, DE remains profoundly enigmatic. Consequently, numerous cosmological models have been proposed and studied over recent years. The simplest approach involves non-interacting models, where DM and DE are conserved independently, resulting in separate evolutionary paths for these components. In contrast, more generalized models allow for interactions between DM and DE, providing a broader framework for understanding their dynamics.

The interaction between DM and DE is a promising mechanism to address the cosmic coincidence problem, although its initial motivation was to resolve the discrepancy in the cosmological constant \cite{delCampo:2008jx,Pavon:2005yx}. Early work by Wetterich demonstrated that an interaction between a scalar field and gravity could result in a dynamic effective cosmological constant that asymptotically approaches a small value, providing a plausible explanation for the mismatched cosmological constant \cite{Wetterich:1994bg}. These dual motivations laid the foundation for exploring interactions within the dark sector. Initially, the primary motivation for interacting dark energy (IDE) models was to address or mitigate the coincidence problem. However, more recently, the focus has shifted toward resolving the discrepancy between the Hubble constant values derived from CMB and local measurements. As the tension between high-redshift and low-redshift Hubble constant measurements has persisted with improving data, IDE models have emerged as promising candidates to reconcile these discrepancies with the $\Lambda$CDM model \cite{Vagnozzi:2017ovm,Aljaf:2020eqh,Cheng:2019bkh,Yang:2017iew, Nong:2024bkr}.

The interaction within the dark sector is largely a phenomenological concept, as no fundamental principle explicitly dictates its existence. However, from a theoretical perspective, particularly in particle physics, any two matter fields, such as DM and DE fields, can interact. This idea has garnered significant attention within the cosmological community because of its potential implications. Notably, allowing such an interaction can transition the DE equation of state (EoS) from the quintessence regime to the phantom regime, effectively introducing a quintom-like behavior. Phenomenological models are typically formulated by incorporating an energy exchange between dark components into their continuity equations with an interacting kernel $\mathcal{U}$. Similarly to how interactions
behave in particle physics, one would expect the kernel to be a function of the energy densities, involved $\rho_{DE}$, $\rho_{CDM}$ and of time, $H^{-1}$. Moreover, within the framework of field theory, considering interactions between the dark sectors is both natural and unavoidable. Exploring these interactions could provide valuable insights into the fundamental nature of DM and DE. Various interaction models that have been proposed and tested in the literature are linear models, such as $\mathcal{U} = \alpha_{m} H \rho_{m}$, $\mathcal{U} = \alpha_{DE}H\rho_{DE}$ and $\mathcal{U} = H(\alpha_{m} \rho_{m} + \alpha_{DE}\rho_{DE})$ \cite{Amendola:2006dg,Quartin:2008px,Pavon:2007gt,Wang:2016lxa}. Although only a few non-linear models have been proposed and investigated in the literature \cite{Wang:2024vmw}. 

Another approach to explaining the accelerated expansion of the universe involves dynamical dark-energy models based on modifying gravity over large distances. Examples of such models include $f(R)$ gravity \cite{Santos:2007bs,Sotiriou:2006hs,Capozziello:2007ms,Capozziello:2008qc}, where the Ricci scalar is replaced by a more general function of $R$ as well as scalar-tensor theories \cite{Amendola:1999qq,Bartolo:1999sq,Nojiri:2015fra,Baker:2020apq}, Galileon gravity \cite{Nicolis:2008in}, and Gauss-Bonnet gravity \cite{Nojiri:2005vv,Nojiri:2005jg}. There are other different modified theories of gravity such as $f(R,T)$ gravity \cite{Harko:2011kv,Alvarenga:2013syu,Shabani:2014xvi,Fisher:2019ekh,Odintsov:2013iba,BarrientosO:2014mys}, $f(\mathcal{T})$ gravity \cite{Cai:2015emx,Paliathanasis:2016vsw,Li:2011wu,Capozziello:2011hj}, where $\mathcal{T}$ is a torsion, $f(Q)$
gravity \cite{Heisenberg:2023lru,J1,Mandal:2020buf,Gadbail_2024,Lazkoz:2019sjl,Frusciante:2021sio,Sokoliuk:2023ccw}, where $Q$ is the nonmetricity. For several decades, dynamical system analysis has been used in cosmology to qualitatively study these models, proving to be effective in identifying and classifying their asymptotic behaviors \cite{Wainwright:2004cd,Coley:1999uh}. Studying autonomous dynamical analysis and point stability is essential for understanding the evolution of the universe and the behavior of cosmological models. Autonomous dynamical systems reformulate the equations governing cosmic evolution into first-order differential equations, simplifying the analysis and enabling a deeper exploration of the universe's dynamics. Critical points within these systems represent key asymptotic behaviors of the universe, such as matter domination, radiation domination, or accelerated expansion. Stability analysis of these points is critical for determining whether the universe can evolve toward specific states, such as the current dark-energy-dominated epoch, thereby linking theoretical predictions with observational data. Moreover, dynamical system analysis provides insights into the influence of model parameters, such as coupling constants or power law indices, on the universe's trajectory, helping identify ranges that lead to physically viable solutions, including late-time acceleration or scaling behaviors. This framework has been applied to investigate various DE models in various modified theories of gravity \cite{BeltranJimenez:2019tme,Bahamonde1,Samart1,Khyllep:2021wjd,Hussain:2024yee,An:2015mvw,Carloni:2015lsa,DAmbrosio:2021pnd,Arora:2022dti}.

The purpose of this work is to investigate the interaction between DE and DM within viable models of $f(Q)$ gravity \cite{BeltranJimenez:2017tkd,BeltranJimenez:2019tme}. This theory extends the alternative to General Relativity (GR) known as the symmetric teleparallel theory, which is based on the nonmetricity scalar $Q$, with both curvature and torsion absent. Geometrically, the nonmetricity $Q$ describes the variation in the length of a vector during parallel transport. Several studies have been done and explored in $f(Q)$ gravity which presents intriguing applications. For review, one can check the references \cite{Harko:2018gxr,Ayuso:2020dcu,Anagnostopoulos:2021ydo,Gadbail:2022jco,Khyllep:2022spx,Gadbail:2023fjh,Arora:2022mlo,Barros:2020bgg,Atayde:2023aoj,Nunes:2018xbm}. The structure of this work is organized as follows: In Section \ref{section 2}, we present the fundamental cosmological equations of the general $f(Q)$ theory and derive the Friedmann equations corresponding to the FLRW metric. In Section \ref{section 3}, we derive the autonomous dynamical system within the framework of $f(Q)$ gravity theory, focusing on the interaction between DE and DM. In Section \ref{section 4}, we apply the power-law $f(Q)$ model to enclose the dynamical system and conduct a further study of $f(Q)$ gravity. Finally, we discuss our findings in Section \ref{section 5}.



\section{\NoCaseChange{Formulation of $f(Q)$ gravity and FLRW cosmology}}
\label{section 2}
In this section, we briefly discuss the formulation of $f(Q)$ gravity, a simple generalization of symmetric teleparallel gravity theory. This theory requires a different curvature and torsion-free connection, i.e., it is wholly dependent on nonmetricity. This formulation introduces new dynamics compared to GR by allowing the function $f(Q)$ to dictate how nonmetricity influences the gravitational interaction. The non-metricity tensor $Q_{\sigma\mu\nu}$ is defined as $Q_{\sigma\mu\nu}=\nabla_{\sigma}g_{\mu\nu}$, which geometrically describes the variation of the length of a vector in the parallel transport.\\
We begin by recalling that the general affine connection allows for a decomposition \cite{Nester:1998mp}, which can be systematically expressed as follows:
\begin{equation}
\hat{\Gamma}^{\,\sigma}_{\,\,\,\mu\nu}=\Gamma^{\,\sigma}_{\,\,\,\mu\nu}+K^{\,\sigma}_{\,\,\,\mu\nu}+L^{\,\sigma}_{\,\,\,\mu\nu},
\end{equation}
where the Levi-Civita connection $\Gamma^{\,\sigma}_{\,\,\,\mu\nu}$ is defined as
\begin{equation}
\Gamma^{\,\sigma}_{\,\,\,\mu\nu} =\frac{1}{2}g^{\sigma\lambda}\left(\partial_{\mu}g_{\lambda\nu}+\partial_{\nu}g_{\lambda\mu}-\partial_{\lambda}g_{\mu\nu}\right),
\end{equation}
which can be uniquely determined by the first-order derivatives of the metric tensor $g_{\mu\nu}$. The contortion $K^{\,\sigma}_{\,\,\,\mu\nu}$ and deformation tensor $L^{\,\sigma}_{\,\,\,\mu\nu}$ are defined as
\begin{eqnarray*}
K^{\,\sigma}_{\,\,\,\mu\nu} &=& \frac{1}{2}T^{\,\sigma}_{\,\,\,\mu\nu}+T^{\,\,\,\,\,\,\,\sigma}_{(\mu\,\,\,\,\,\,\nu)},\\
L^{\,\sigma}_{\,\,\,\mu\nu} &=& -\frac{1}{2}g^{\sigma\lambda}\left(Q_{\mu\lambda\nu}+Q_{\nu\lambda\mu}-Q_{\lambda\mu\nu}\right),
\end{eqnarray*}
respectively, which describes non-Riemannian properties in the manifold. The contortion tensor disappears in the symmetric teleparallel theory because it follows an anti-symmetric property. The interplay between nonmetricity and the absence of torsion would influence cosmological models and the evolution of the universe. These effects could manifest in scenarios such as the dynamics of inflation, the behavior of DE, and the formation of large-scale structures.\\
With the aid of the nonmetricity tensor, we can define the superpotential tensor $P_{\,\,\mu\nu}^{\sigma}$ as
\begin{equation}
\label{2}
4P_{\,\,\mu\nu}^{\sigma}=-Q^{\sigma}_{\,\,\,\,\mu\nu}+2Q^{\,\,\,\,\,\,\sigma}_{(\mu\,\,\,\,\nu)}-Q^{\sigma}g_{\mu\nu}-\tilde{Q}^{\sigma}g_{\mu\nu}-\delta^{\sigma}_{(\mu}\, Q\,_{\nu)},
\end{equation} 
where $Q_{\sigma}=Q_{\sigma\,\,\,\,\mu}^{\,\,\,\,\mu}$ and $\tilde{Q}_{\sigma}=Q^{\mu}_{\,\,\,\,\sigma\mu}\,$. From the above quantities, we can obtain a non-metricity scalar as 
\begin{equation}
\label{3}
Q=-Q_{\sigma\mu\nu}P^{\sigma\mu\nu}.
\end{equation}
The action for $f(Q)$ gravity \cite{BeltranJimenez:2019tme}
\begin{equation}
\label{4}
S=\int \left[-\frac{1}{2\kappa^2}f(Q)+\mathcal{L}_m\right]\sqrt{-g}\,d^4x,
\end{equation}
where $f(Q)$ represents any function of the scalar $Q$, $g$ denotes the determinant of $g_{\mu\nu}$, and $\mathcal{L}_m$ stands for the matter Lagrangian density. Here \(\kappa^2 = 8\pi G\), where \(G\) is the Newtonian gravitational constant.\\
The equations of motion in $f(Q)$ gravity are derived by varying the action with respect to the metric. For simplicity, we set $8\pi G = 1 $. This leads to a set of modified field equations that incorporate the effects of non-metricity, providing a richer structure for modeling gravitational phenomena, and is written as
\begin{eqnarray}
\label{GFE}
\frac{2}{\sqrt{-g}}\nabla_{\sigma}\left(f_Q\sqrt{-g}\,P^{\sigma}_{\,\,\mu\nu}\right)+\frac{1}{2}f(Q)\,g_{\mu\nu}
+f_Q\left(P_{\mu\sigma\lambda}Q_{\nu}^{\,\,\,\sigma\lambda}-2Q_{\sigma\lambda\mu}P^{\sigma\lambda}_{\,\,\,\,\,\,\nu}\right)=T_{\mu\nu},
\end{eqnarray}
where $f_Q=\frac{d f}{d Q}$. The energy-momentum tensor for matter is now defined as
$T_{\mu\nu}\equiv-\frac{2}{\sqrt{-g}}\frac{\delta(\sqrt{-g})\mathcal{L}_m} {\delta g^{\mu\nu}}$.\\
Also, by varying action \eqref{5} with respect to the connection results in
\begin{equation}
\label{conn}
    \nabla_{\sigma}\lambda_{k}^{\,\,\,\mu\nu\sigma}+\lambda_{k}^{\,\,\,\mu\nu}=\sqrt{-g}f_Q\,P_{\,k}^{\,\,\,\mu\nu}+H_{\,k}^{\,\,\,\mu\nu},
\end{equation}
where $H_{\,k}^{\,\,\,\mu\nu}=-\frac{1}{2}\frac{\delta\mathcal{L}_m}{\delta\Gamma^{k}_{\,\,\,\mu\nu}}$ is the hypermomentum tensor density.\\

It is possible to simplify Eq. \eqref{conn} by taking into account the antisymmetry property of $\mu$ and $\nu$ in the Lagrangian multiplier coefficients
\begin{equation}
    \nabla_{\mu}\nabla_{\nu} \left(f_{Q}\sqrt{-g}\,P_{\,k}^{\,\,\,\mu\nu}+H_{\,k}^{\,\,\,\mu\nu}\right)=0.
\end{equation}
More specifically, the connection can be parameterized with a collection of functions $\xi^{\alpha}$ as
$\hat{\Gamma}^{\,\sigma}_{\,\,\,\alpha\beta}=\frac{\partial x^{\sigma}}{\partial\xi^{\mu}}\partial_{\alpha}\partial_{\beta}\xi^{\mu}$. The connection equation of motion can be easily calculated by noticing that the variation of the connection with respect to $\xi^{\sigma}$ is equivalent to performing a diffeomorphism so that $\delta_{\xi}\hat{\Gamma}^{\,\sigma}_{\,\,\,\mu\nu}=-\mathcal{L}_{\xi}\hat{\Gamma}^{\,\sigma}_{\,\,\,\mu\nu}=-\nabla_{\mu}\nabla_{\nu}\xi^{\sigma}$, where we have used that the connection is flat and torsion-free \cite{BeltranJimenez:2019tme}. Furthermore, in the absence of hypermomentum,
the connection field equations read as
\begin{equation}
\label{8}
\nabla_{\mu}\nabla_{\nu} \left(f_{Q}\sqrt{-g}\,P_{\,k}^{\,\,\,\mu\nu}\right)=0.
\end{equation}
Symmetric teleparallel gravity provides a geometric formulation of gravity that is fully equivalent to general relativity. This equivalence becomes apparent in the coincident gauge, where the connection satisfies $\hat{\Gamma}^{\,\sigma}_{\,\,\,\mu\nu} = 0$. Imposing the condition that the connection is symmetric eliminates the torsion tensor, allowing the Levi-Civita connection to be expressed in terms of the disformation tensor as $\hat{\Gamma}^{\,\sigma}_{\,\,\,\mu\nu} = -L^{\,\sigma}_{\,\,\,\mu\nu}$. Consequently, the non-metricity simplifies to $Q_{\sigma\mu\nu} = \partial_{\sigma} g_{\mu\nu}$. The use of the coincident gauge in \( f(Q) \) gravity offers significant advantages, such as simplifying calculations by eliminating the connection, reducing ambiguities, and isolating the effects of non-metricity. This gauge allows for more tractable analytical and numerical analyses, making it particularly useful for cosmological studies. However, it comes with limitations, including a loss of generality, as results are gauge-dependent and may not fully capture the broader implications of the connection in \( f(Q) \) gravity. Additionally, the coincident gauge may introduce biases and restrict the exploration of the connection's physical role and the general behavior of the non-metricity scalar \( Q \).\\

To apply $ f(Q)$ gravity in a cosmological context, we consider the spatially flat Friedmann-Lema\^{i}tre-Robertson-Walker (FLRW) spacetime, characterized by the metric

\begin{equation}
\label{6}
ds^2=-dt^2+a^2(t)\,\delta_{ij}\,dx^i\,dx^j,\,\,\,\,\,\,(i,j=1,2,3),
\end{equation}
where $a(t)$ is the cosmological scale factor. For this metric, the corresponding non-metricity scalar is given by  $Q = 6H^2$, with $ H = \frac{\dot{a}}{a}$ representing the Hubble parameter, and the dot denotes a derivative with respect to the coordinate time $t$. Applying the FLRW metric to the general field equation \eqref{GFE}, the Friedman equations of $f(Q)$ cosmology read as  
\begin{eqnarray}
\label{f1}
6H^2f_Q-\frac{f}{2}&=&\rho\,,\\
\label{f2}
\left(12H^2\,f_{QQ}+f_Q\right)\dot{H}&=&-\frac{1}{2}(p+\rho),
\end{eqnarray}
where $\rho$ and $p$ are energy density and pressure for the perfect fluid, respectively, while $f_Q=\frac{df}{dQ}$, and $f_{QQ}=\frac{d^2f}{dQ^2}$.\\
To study the interaction between DE and DM, we have to modify the conservation equations of the DE and the DM given above by adding some coupling term $\mathcal{U}$ as \cite{Samart1}
 \begin{eqnarray*}
  && \dot{\rho}_{DE} + 3H(p_{DE}+\rho_{DE})=\mathcal{U}\,, \\
  && \dot{\rho}_{m} + 3H\,\rho_m =-\mathcal{U}\,,\\
  && \dot{\rho}_{r}+ 4H\,\rho_r =0\,.
\end{eqnarray*}
In addition, the coupling term $\mathcal{U}$ between DE and DM can be interpreted as the exchange rate of energy density between these two components. When \(\mathcal{U} > 0\), energy is transferred from DM to DE, whereas when \(\mathcal{U} < 0\), energy is transferred from DE to DM.

\section{\NoCaseChange{Autonomous dynamical system of interacting DE
and DM}}
\label{section 3}
In this section, we derive the autonomous dynamical system within the framework of $f(Q)$ gravity theory, focusing on the interaction between DE and DM. Our approach involves incorporating two interacting terms. The first term relies solely on the energy density of the matter sector, the Hubble parameter, and a coupling constant $\alpha$, structured multiplicatively as $\mathcal{U}=3\alpha\,H\,\rho_m$. The second term encompasses the multiplication of energy densities from both sectors, capturing the immediate effects of both DM and DE on the interaction term, with a dimensionless coupling constant $\alpha$, denoted as $\mathcal{U}=\frac{\alpha}{3H}\rho_m\,\rho_{DE}$.\\
According to the Friedmann equation \eqref{f1}, we can define the dimensionless variable as
\begin{equation}
\label{16}
    x=\frac{f}{12H^2\,f_Q},\,\,\,\,\,y=\frac{\rho_r}{6H^2\,f_Q}.
\end{equation}
Invoking the above dimensionless variables, it is straightforward to derive the autonomous equations which play a key role for studying the dynamical system for the interacting DE and DM.

\subsection{Case I : $\mathcal{U}=3\alpha\,H\,\rho_m$}

Using Eq.\eqref{f1}, the autonomous dynamical system is given by
\begin{equation}
    \frac{dx}{dN}=(1-x)\frac{\dot{H}}{H^2}+3x-3x^2+xy+3\alpha\,x(1-x-y),
     \end{equation}
\begin{equation}
    \frac{dy}{dN}=-\frac{\dot{H}}{H^2}\,y-y+y^2-3xy+3\alpha\,y(1-x-y).
\end{equation}
where the variable $N =ln\,a$ is e-folding number and leads to $\frac{d}{dN}=\frac{1}{H}\frac{d}{dt}$.

\subsection{Case II : $\mathcal{U}=\frac{\alpha}{3H}\rho_m\,\rho_{DE}$}
For this particular case, the autonomous dynamical system is given by
\begin{equation}
    \frac{dx}{dN}=(1-x)\frac{\dot{H}}{H^2}+3x-3x^2+xy+\alpha\,x^2(1-x-y),
 \end{equation}
\begin{equation}
    \frac{dy}{dN}=-\frac{\dot{H}}{H^2}\,y-y+y^2-3xy+\alpha\,xy(1-x-y).
\end{equation}
Furthermore, we obtain the generalized equation for $\frac{\dot{H}}{H^{2}}$ as 
\begin{equation}
\label{21}
    \frac{\dot{H}}{H^{2}} = \frac{-3f_{Q}(1-x+\frac{y}{3})}{2Qf_{QQ}+f_{Q}}.
\end{equation}
Consequently, the density parameters for individual matter species can be linked to the dimensionless variables outlined in Eq. \eqref{16}. These variables serve as a set of constraint equations:
\begin{equation}
    \Omega_m=1-x-y,\,\,\,\,\,\,\,\Omega_r=y,\,\,\,\,\,\,\,\Omega_{DE}=x.
\end{equation}
Finally, we define the EoS and deceleration parameters corresponding to the dimensionless variables to check the appropriate acceleration expansion of the universe, that is
\begin{equation}
    \omega=-1+\frac{2f_{Q}(1-x+\frac{y}{3})}{2Qf_{QQ}+f_{Q}}
\end{equation}
and 
\begin{equation}
    q=-1+\frac{3f_{Q}(1-x+\frac{y}{3})}{2Qf_{QQ}+f_{Q}}.
\end{equation}
Both $\omega$ and $q$ represent different aspects of cosmic evolution, each uniquely influencing the large-scale structure of the universe. Grasping the significance of these parameters is essential for studying the effects of DE across various stages of cosmic evolution. Moreover, these parameters aid in comparing and differentiating between various DE models, each employing distinct mechanisms to drive cosmic acceleration.


\section{Model:- $f(Q)=6\gamma\,H_0^2\left(\frac{Q}{Q_0}\right)^n$}
\label{section 4}
In the previous section, we derived autonomous dynamical systems, and our primary objective is to investigate and analyze them. To accomplish this goal, we will focus on the designated power-law $f(Q)$ model, which has the form $f(Q)=6\gamma\,H_0^2\left(\frac{Q}{Q_0}\right)^n$, where $\gamma$, $n$ are free model parameters and $Q_{0}=6H_0^2$.\\
Corresponding to our power-law $f(Q)$ model, Eq. \eqref{21} can be simplified to obtain
\begin{equation}
    \frac{\dot{H}}{H^2}=-\frac{3(1-x+\frac{1}{3}y)}{2n-1}.
\end{equation}
We will now derive the critical points of the system by setting $dx/dN = dy/dN = 0$ and determine the corresponding eigenvalues for both dynamical systems. The critical points, which represent the solutions of the dynamical system, provide an initial qualitative understanding of the phase space. As discussed below, these points can be classified based on their stability properties. In the absence of singularities or strange attractors, the trajectories of $x(N)$ and $y(N)$, which are generally obtained numerically, tend to evolve from unstable fixed points to stable fixed points, possibly passing through intermediate saddle points.\\ 
Analyzing the critical points and evaluating their stability is essential for a thorough understanding of the critical aspects of cosmic evolution driven by interactions between DE and DM, as explored in this study. Furthermore, the properties of the dynamical system depend significantly on the values of the constants $\alpha$ and $n$.

\subsection{For case I}
The fixed points $(x,y)$ of the general dynamical system are described in table \ref{TABLE-I}.
Let us now focus on the dynamics of each critical point and its features in the following subsections.

\begin{table*}[!htb]
\centering 
\begin{tabular}{|*{6}{c|}}\hline 
    \parbox[c][0.8cm]{3.3cm}{Critical Points} & $\Omega_m$ & $\Omega_{DE}$ & $w$ & Eigenvalues & Stability\\ [0.5ex]\hline \hline 
    \parbox[c][1.2cm]{3cm}{$P_1:(1,0)$} & $0$ & $1$ & $-1$ & $\{ -4,\,\,\,-3 (\alpha +1) \}$ & \begin{tabular}{@{}c@{}}Stable node for $\alpha>-1$, \\ Saddle for $\alpha<-1$,\\  Non-hyperbolic for $\alpha=-1$.\end{tabular}\\
    \hline
    \parbox[c][1cm]{3cm}{$P_2:\left(\frac{1}{2n},\frac{2n-1}{2n}\right)$} & $0$ & $\frac{1}{2n}$ & $-1+\frac{4}{3n}$ & $\{4,1-3\alpha\}$  &  \begin{tabular}{@{}c@{}}Saddle node for $\alpha>1/3$,\\ Unstable for $\alpha<1/3$,\\ Non-hyperbolic for $\alpha=1/3$.
    \end{tabular}\\
    \hline
    \parbox[c][0.8cm]{3cm}{$P_3:\left(\frac{1}{2n+\alpha(2n-1)},0\right)$} & $\frac{1}{\alpha -2 (\alpha +1) n}+1$ & $\frac{1}{2n+\alpha(2n-1)}$ & $\frac{3 \alpha -2 (\alpha +1) n+2}{2 (\alpha +1) n-\alpha }$ & $\{3 (\alpha +1),3 \alpha -1\}$ & \begin{tabular}{@{}c@{}}Stable node for $\alpha<-1$,\\ Saddle for $-1<\alpha<1/3$,\\ Unstable node for $\alpha>1/3$.
    \end{tabular} \\
    \hline
\end{tabular}
\caption{This table summarizes the analysis of the critical points for Case I. }
\label{TABLE-I}
\end{table*}

\subsubsection{$P_1$: the de Sitter fixed point}
We start the discussion with the first fixed point. Here we have
\begin{equation*}
    P_1: (x,y)=(1,0).
\end{equation*}
This critical point is independent of both the model and coupling parameters. Corresponding to this fixed point, the matter density, DE density, and effective EoS parameters are
\begin{equation}
   \Omega_{m}=0,\,\,\,\,\Omega_{DE}=1,\,\,\,\,\omega=-1.
\end{equation}
At this point, both DM and radiation are absent, and the universe is dominated by DE. The EoS at this fixed point indicates that the universe is experiencing an accelerated expansion. To study the stability of this fixed point, we can analyze the eigenvalues of the Jacobian matrix, which describe the behavior of the fixed point. The eigenvalues for the fixed point $P_1$ read $  \lbrace -4,\,\,\,-3 (\alpha +1) \rbrace$.
The stability behavior of this fixed point (which depends on $\alpha$) is obtained as
\begin{itemize}
    \item Stable node for $\alpha>-1$,
    \item Saddle for $\alpha<-1$,
    \item Non-hyperbolic for $\alpha=-1$.
\end{itemize}

\subsubsection{$P_2$: Non-metricity dominated fixed point}
Next, for the second fixed point, we have
\begin{equation*}
    P_2: (x,y)=\left(\frac{1}{2n},\frac{2n-1}{2n}\right).
\end{equation*}
The characteristics of this critical point solely depend on the model parameter $n$.
 Corresponding to this fixed point, the matter density, DE density, and effective EoS parameters are obtained as
\begin{equation}
   \Omega_{m}=0,\,\,\,\,\Omega_{DE}=\frac{1}{2n},\,\,\,\,\omega=-1+\frac{4}{3n}.
\end{equation}
The conditions for an accelerating universe are $n<0$ or $n>2$. When $n<0$, the universe exhibits phantom-like behavior, i.e., $\omega<-1$. When $n>2$, the universe exhibits quintessence-like behavior, i.e., $-1<\omega<-1/3$. In this fixed-point solution, a matter-dominated era occurs for $n=4/3$ and a radiation-dominated era for $n=1$. The eigenvalues for the fixed point $P_2$ are as follows $ \lbrace 4,\,\,\,1-3\alpha \rbrace$. The stability behavior for this particular fixed point can be obtained for different $\alpha$, given by
\begin{itemize}
    \item Saddle node for $\alpha>1/3$,
    \item Unstable for $\alpha<1/3$,
    \item Non-hyperbolic for $\alpha=1/3$.
\end{itemize}

\subsubsection{$P_3$: scaling solution fixed point}

\begin{figure}[]
\centering
\includegraphics[scale=0.55]{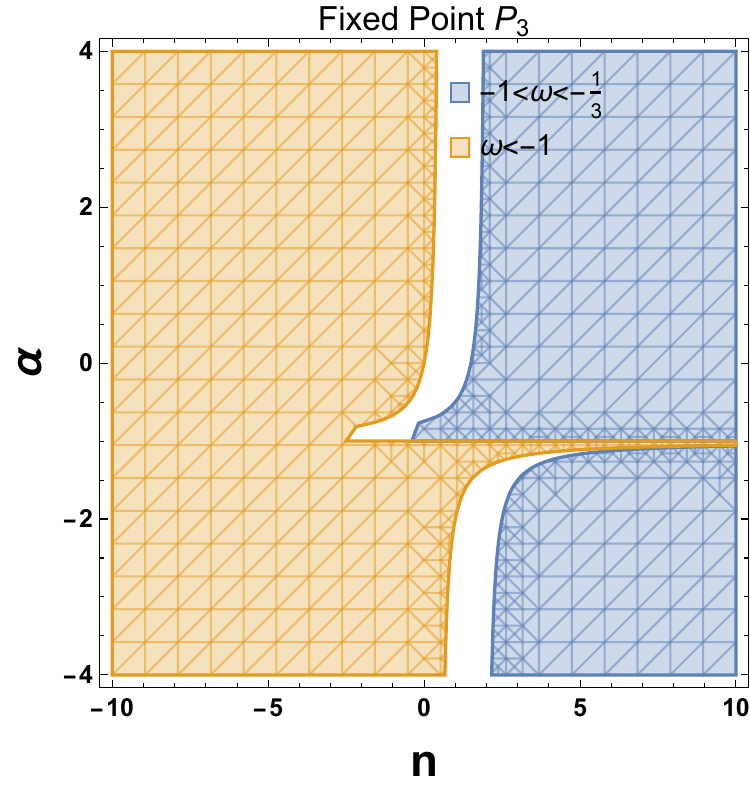}
\caption{\justifying The plot illustrates the relationship between the model parameter $n$ and the interaction parameter $\alpha$ for an accelerating universe.  The light blue region represents the quintessence-like behavior of an accelerating universe (i.e., $-1<\omega<-1/3$), while the golden region represents the phantom-like behavior of an accelerating universe (i.e., $\omega<-1$).} 
\label{1}
\end{figure}

It is worth noting that our result modifies the scaling solution of $f(Q)$ gravity with the interacting DE parameter $\alpha$. The fixed point $P_3$ reads
\begin{equation*}
     P_3: (x,y)=\left(\frac{1}{2n+\alpha(2n-1)},0\right).
\end{equation*}
The properties of this critical point are influenced by both the model parameter $n$ and the coupling parameter $\alpha$. Corresponding to this fixed point, the matter density, DE density, and effective EoS parameters read as
\begin{equation*}
   \Omega_{m}=\frac{1}{\alpha -2 (\alpha +1) n}+1,\quad \Omega_{DE}=\frac{1}{2n+\alpha(2n-1)}, \quad \omega=\frac{3 \alpha -2 (\alpha +1) n+2}{2 (\alpha +1) n-\alpha }.
\end{equation*}
The conditions for an accelerating universe are shown in Fig. \ref{1}.
In this fixed point solution, we obtained a matter-dominated era for $n=\frac{3 \alpha +2}{2 (\alpha +1)}$, $1+\alpha\neq 0$, and radiation-dominated era for $n=\frac{5 \alpha +3}{4 (\alpha +1)}$, $1+\alpha\neq 0$. The eigenvalues for the fixed point $P_3$ are $\lbrace 3 (\alpha +1),\,\,3 \alpha -1 \rbrace$.
We determine the stability of this fixed point for various ranges of $\alpha$. The ranges are as follows
\begin{itemize}
    \item Stable node for $\alpha<-1$,
    \item Saddle for $-1<\alpha<1/3$,
    \item Unstable node for $\alpha>1/3$.
\end{itemize}

In Figs. \ref{2} and \ref{3}, we illustrate the evolution of \(\Omega_m\), \(\Omega_r\), \(\Omega_{DE}\), \(q\), and \(\omega\) for different values of the model parameter \(n\) and the interaction parameter \(\alpha\). This is done using the numerical solution of the dynamical system with the initial conditions $x(0)=0.7$ and $y(0)=0.00005$. The motivation behind the choice of initial conditions is as follows: \( x \) represents the DE density, which currently has a value of $0.7$. Similarly, \( y \) represents the radiation density, which currently has a value of $0.00005$. 
\begin{widetext}

\begin{figure}[]
\centering
\includegraphics[scale=0.65]{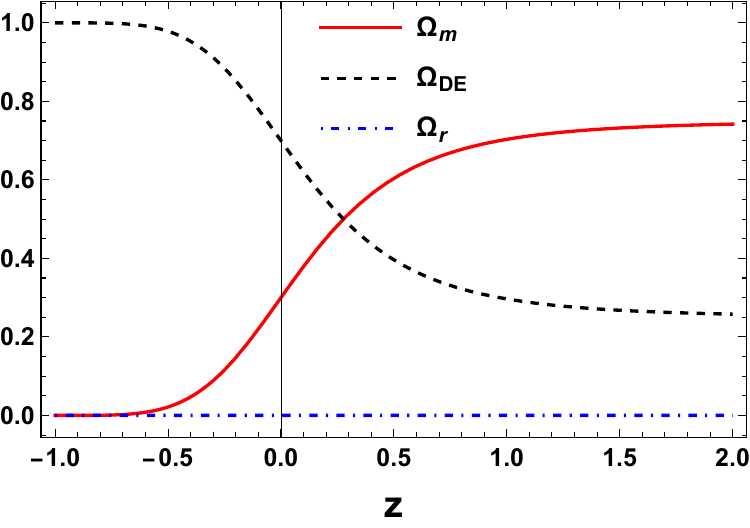} \hspace{0.2in}
\includegraphics[scale=0.67]{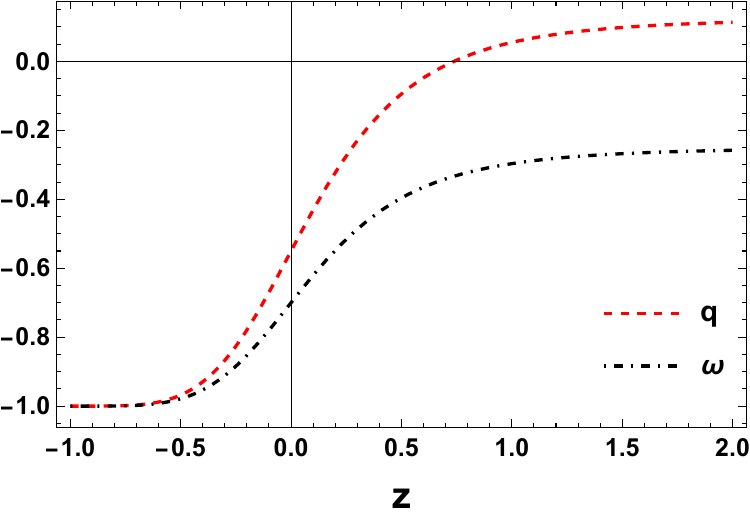}
\caption{\justifying The left panel shows the density parameters for matter ($\Omega_m$), DE ($\Omega_{DE}$), and radiation ($\Omega_r$) as functions of redshift ($z$). The right panel displays the deceleration parameter ($q$) and the equation of state parameter ($\omega$) as functions of redshift ($z$). Together, these two panels illustrate the trajectory for a positive coupling parameter, which can facilitate the transfer of energy from DM to DE.}
\label{2}
\end{figure}

\begin{figure}[]
\centering
\includegraphics[scale=0.65]{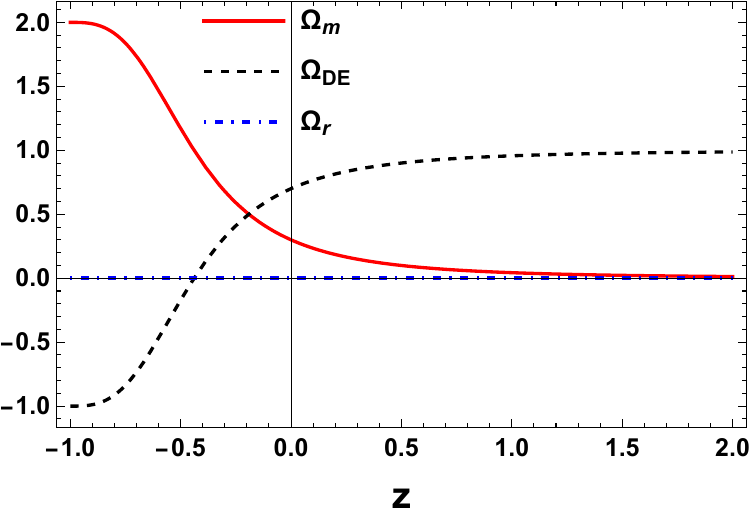} \hspace{0.2in}
\includegraphics[scale=0.65]{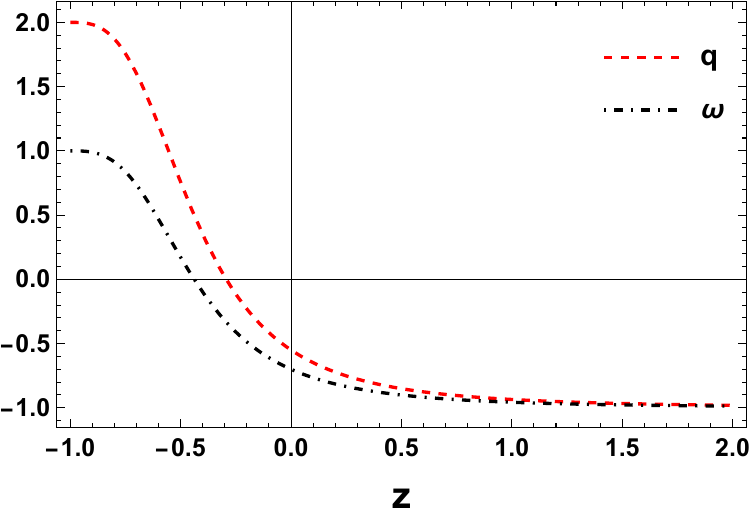}
\caption{\justifying The left panel displays density parameters for matter ($\Omega_m$), DE ($\Omega_{DE}$), and radiation ($\Omega_r$) as functions of redshift $z$. The right panel shows the deceleration parameter $q$ and the EoS parameter $\omega$ as functions of redshift $z$. These two panels illustrate the trajectory for the negative coupling parameter, which can transfer energy from DE to DM.} 
\label{3}
\end{figure}

\end{widetext}
For \(n = 3/2\) and \(\alpha = 1/2\) in Fig. \ref{2}, \(\alpha > 0\) indicates that the coupling term \(\mathcal{U} > 0\), signifying energy transfer from DM to DE. This figure shows that the universe was dominated by matter in the early stages and will be dominated by DE in later stages. Currently, the universe is dominated by DE, with parameters value \(\Omega_m = 0.3\), \(\Omega_r = 0.00005\), \(\Omega_{DE} = 0.7\), \(q_0 = -0.55\), and \(\omega_0 = -0.70\). For these values, fixed point \(P_1\) is stable and represents the de Sitter acceleration solution, while fixed point \(P_2\) is a saddle-node, and \(P_3\) is an unstable node that cannot demonstrate universal acceleration.\\
In Fig. \ref{3}, with \(n = 3/2\) and \(\alpha = -2\), \(\alpha < 0\) indicates that the coupling term \(\mathcal{U} < 0\), that means energy transfers from DE to DM. This figure shows that the universe was dominated by DE in the early stages and will be dominated by DM at later stages. Currently, the universe remains dominated by DE, with parameters value \(\Omega_m = 0.3\), \(\Omega_r = 0.00005\), \(\Omega_{DE} = 0.7\), \(q_0 = -0.563\), and \(\omega_0 = -0.71\). Here, fixed point \(P_3\) is stable and exhibits acceleration for the early and present universe but fails to show acceleration for late times, while fixed point \(P_1\) is a saddle-node, and \(P_2\) is an unstable node. Fig. \ref{4} shows the phase space trajectories of the fixed points \(P_1\), \(P_2\), and \(P_3\).

\begin{figure}[]
\centering
\includegraphics[scale=0.5]{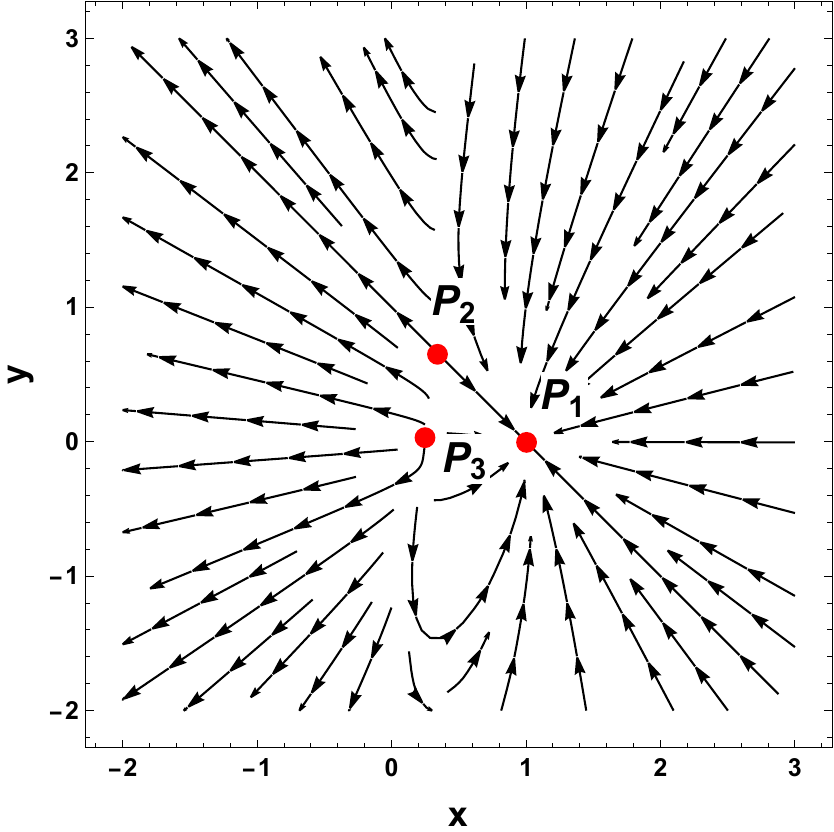} \hspace{0.2in} 
\includegraphics[scale=0.5]{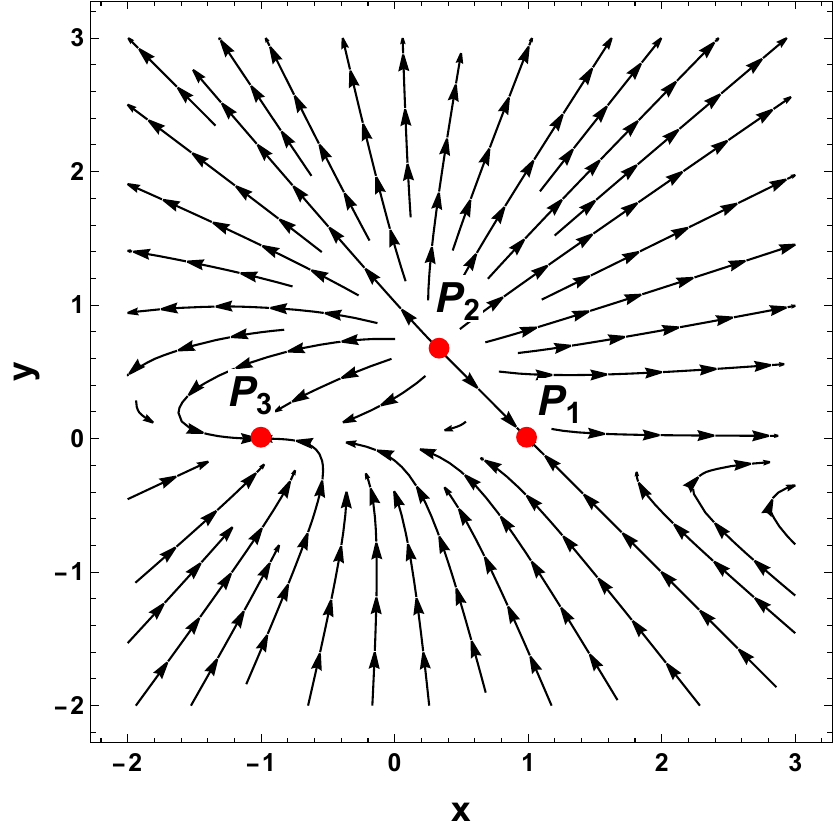}
\caption{\justifying Phase space of $x$ vs $y$. The left phase plot illustrates the stable behavior of $P_1$ when $\alpha>0$, while the right phase plot showcases the stable behavior of $P_3$ for $\alpha<-1$.} 
\label{4}
\end{figure}

\subsection{For case II}
Next, the second case in $f(Q)$ gravity is also investigated in detail for the interacting DE system. The models have been systematically constructed with possible ranges of their model parameters derived from the dynamical system analysis within the interacting DE framework. It is important to examine how the interaction affects the cosmological viability of the $f(Q)$ gravity models.

The fixed points $(x,y)$ of the dynamical system are described in table \ref{TABLE-II}.
\begin{table}[!htb]
\centering 
\begin{tabular}{|*{6}{c|}}\hline 
    \parbox[c][0.8cm]{3.3cm}{Critical Points} & $\Omega_m$ & $\Omega_{DE}$ & $w$ & Eigenvalues & Stability\\ [0.5ex]\hline \hline 
    \parbox[c][1.2cm]{3cm}{$R_1:(1,0)$} & $0$ & $1$ & $-1$ & $\{-4,-3 -\alpha\}$ & \begin{tabular}{@{}c@{}}Stable node for $\alpha>-3$,\\ Saddle for $\alpha<-3$,\\ Non-hyperbolic for $\alpha=-3$.\end{tabular}\\
    \hline
    \parbox[c][1cm]{3cm}{$R_2:\left(\frac{1}{2n},\frac{2n-1}{2n}\right)$} & $0$ & $\frac{1}{2n}$ & $-1+\frac{4}{3n}$ & $\{4,1-\frac{\alpha}{2n}\}$  &  \begin{tabular}{@{}c@{}}Saddle node for $\alpha>2n$,\\ Unstable for $\alpha<2n$,\\ Non-hyperbolic for $\alpha=2n$.
    \end{tabular}\\
    \hline
    \parbox[c][0.8cm]{3cm}{$R_3:\left(\frac{1}{\alpha },\frac{-3 \alpha +8 n-1}{\alpha }\right)$} & $4-\frac{8 n}{\alpha }$ & $\frac{1}{\alpha }$ & $\frac{8}{3 \alpha }-1$ & $\{\mathcal{A}_1-\frac{4n}{\alpha}+4,-\mathcal{A}_1-\frac{4n}{\alpha}+4\}$ & \begin{tabular}{@{}c@{}}The stability conditions\\ are shown in Fig. \ref{5}
     \end{tabular} \\
     \hline
     \parbox[c][2cm]{3cm}{$R_4:\left(\frac{-\sqrt{-3\alpha +9 n^2+6 \alpha  n}-3 n}{2\alpha n-\alpha },0\right)$} & Sec.\ref{R4}    &  \ref{R4} &  \ref{R4} & $\left\lbrace  \frac{5n-1+\mathcal{A}_2}{1-2n},\frac{-2\mathcal{A}_2^2-2(3n-\alpha+2n\alpha)\mathcal{A}_2}{(2n-1)^2\alpha}\right\rbrace$ & \begin{tabular}{@{}c@{}}The stability conditions\\ are shown in Fig. \ref{5}
     \end{tabular} \\
     \hline
     \parbox[c][2cm]{3cm}{$R_5:\left(\frac{ \sqrt{-3\alpha +9 n^2+6 \alpha  n}-3 n}{2 \alpha  n-\alpha },0\right)$} & Sec.\ref{R5} & \ref{R5} & \ref{R5} & $\left\lbrace \frac{-2\mathcal{A}_2^2+2(3n-\alpha+2n\alpha)\mathcal{A}_2}{(2n-1)^2\alpha},\frac{1-5n+\mathcal{A}_2}{2n-1}\right\rbrace$ & \begin{tabular}{@{}c@{}}The stability conditions\\ are shown in Fig. \ref{5}
     \end{tabular} \\
     \hline
\end{tabular}
\caption{\justifying This figure summarizes the results of the analysis of the critical points for Case II. The critical points \( R_4 \) and \( R_5 \), however, are difficult to present comprehensively in a table format. Detailed analyses of these points can be found directly in Sections \ref{R4} and \ref{R5}.}
\label{TABLE-II}
\end{table}
The expressions of $\mathcal{A}_1$ and $\mathcal{A}_2$ in the table are given by $\mathcal{A}_1=\frac{2 \sqrt{(2 n-1) \left(3 (\alpha -1) \alpha ^2+8 n^3-4 n^2-6 (\alpha -1) \alpha  n\right)}}{\alpha -2 \alpha  n}$ and $\mathcal{A}_2=\sqrt{9n^2-3\alpha+6n\alpha}$.

\subsubsection{$R_1$: the de Sitter fixed point} \label{sub2.1}

Corresponding to fixed point $R_1: (x,y)=(1,0)$, the matter density, DE density, and effective EoS parameters are
\begin{equation}
   \Omega_{m}=0,\,\,\,\,\Omega_{DE}=1,\,\,\,\,\omega=-1.
\end{equation}
At this point, both DM and radiation are absent, and the universe is dominated by DE. The EoS at this fixed point indicates that the universe is experiencing an accelerated expansion. To study the stability of this fixed point, we can analyze the eigenvalues of the Jacobian matrix, which describe the behavior of the fixed point. The eigenvalues for the fixed point $R_1$ are $\lbrace{ -4,\,\,\,-\alpha -3\rbrace}$.

The stability behavior of this fixed point is as follows: 
\begin{itemize}
    \item Stable node for $\alpha>-3$,
    \item Saddle for $\alpha<-3$,
    \item Non-hyperbolic for $\alpha=-3$.
\end{itemize}
\subsubsection{$R_2$: Non-metricity dominated fixed point}
The fixed point $R_2$ can be used to explain the late-time accelerating expansion of the universe in $f(Q)$ gravity. The point is given by 
\begin{equation*}
    R_2: (x,y)=\left(\frac{1}{2n},\frac{2n-1}{2n}\right)
\end{equation*}
The characteristics of this critical point solely depend on the model parameter $n$. Corresponding to this fixed point, the matter density, DE density, and effective EoS parameters are obtained as
\begin{equation}
   \Omega_{m}=0,\,\,\,\,\Omega_{DE}=\frac{1}{2n},\,\,\,\,\omega=-1+\frac{4}{3n}.
\end{equation}
The conditions for an accelerating universe are $n<0$ or $n>2$. When $n<0$, the universe exhibits phantom-like behavior, i.e., $\omega<-1$. When $n>2$, the universe exhibits quintessence-like behavior, i.e., $-1<\omega<-1/3$. In this fixed-point solution, we have obtained a matter-dominated era for $n=4/3$ and a radiation-dominated era for $n=1$.\\
The corresponding eigenvalues for the fixed point $R_2$ are $\lbrace{ 4,\,\,1-\frac{\alpha}{2n} \rbrace}$. We determine the stability of this fixed point for various ranges of $\alpha$. The ranges are as follows
\begin{itemize}
    \item Saddle node for $\alpha>2n$,
    \item Unstable for $\alpha<2n$,
    \item Non-hyperbolic for $\alpha=2n$.
\end{itemize}

\subsubsection{$R_3$: Scaling solution point}\label{sub2.2}
\begin{figure}[]
\centering
\includegraphics[scale=0.6]{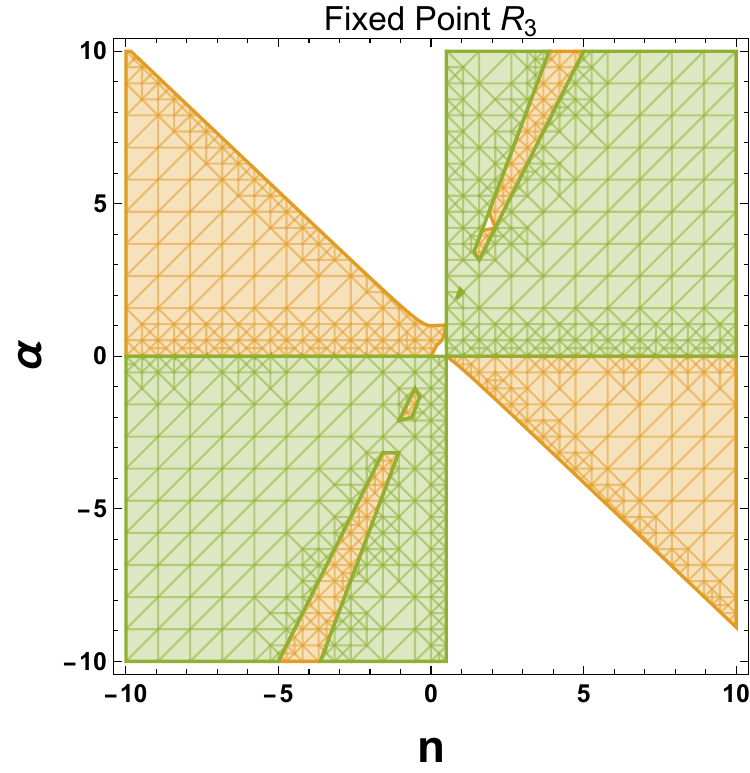}
\caption{\justifying The region plot illustrates the stable, unstable, and saddle behaviors associated with the fixed point $R_3$. In this plot, the stable region is very small $\left(\frac{1}{8}<n\leq 0.4393\,\, \text{and}\,\, 0<\alpha <\frac{1}{3} (8 n-1)\right)$. } 
\label{5}
\end{figure}

The fixed point $R_3$ represents a scaling solution for the universe. This scaling solution makes the ratio $\Omega_{m}/\Omega_{DE}$ constant. The fixed point in this case is given by
\begin{equation*}
    R_3: (x,y)=\left(\frac{1}{\alpha },\frac{-3 \alpha +8 n-1}{\alpha }\right).
\end{equation*}
Corresponding to this fixed point, the matter density, DE density, and effective EoS parameters read as
\begin{equation}
    \Omega_m=4-\frac{8 n}{\alpha },\,\,\,\,\Omega_{DE}=\frac{1}{\alpha },\,\,\,\,\omega=\frac{8}{3 \alpha }-1.
\end{equation} 
The conditions for an accelerating universe are $\alpha<0$ or $\alpha>4$. When $\alpha<0$, the universe exhibits phantom-like behavior, i.e., $\omega<-1$. When $\alpha>4$, the universe exhibits quintessence-like behavior, i.e., $-1<\omega<-1/3$. In this fixed-point solution, we have obtained a matter-dominated era for $\alpha=8/3$ and a radiation-dominated era for $\alpha=2$. The stability conditions are shown in Fig. \ref{5}.



\subsubsection{$R_4$: Scaling solution point}\label{R4}

The fixed point $R_4$ represents the scaling solution point of the universe. In addition, it is worth noting that our result modifies the scaling solution with the interacting DE parameter $\alpha$. The point $R_4$ is 
\begin{equation*}
    R_4: (x,y)=\left(\frac{-\sqrt{-3\alpha +9 n^2+6 \alpha  n}-3 n}{2 \alpha  n-\alpha },0\right).
\end{equation*}

\begin{figure}
     \centering
     \begin{subfigure}[b]{0.45\textwidth}
         \centering
         \includegraphics[width=\textwidth]{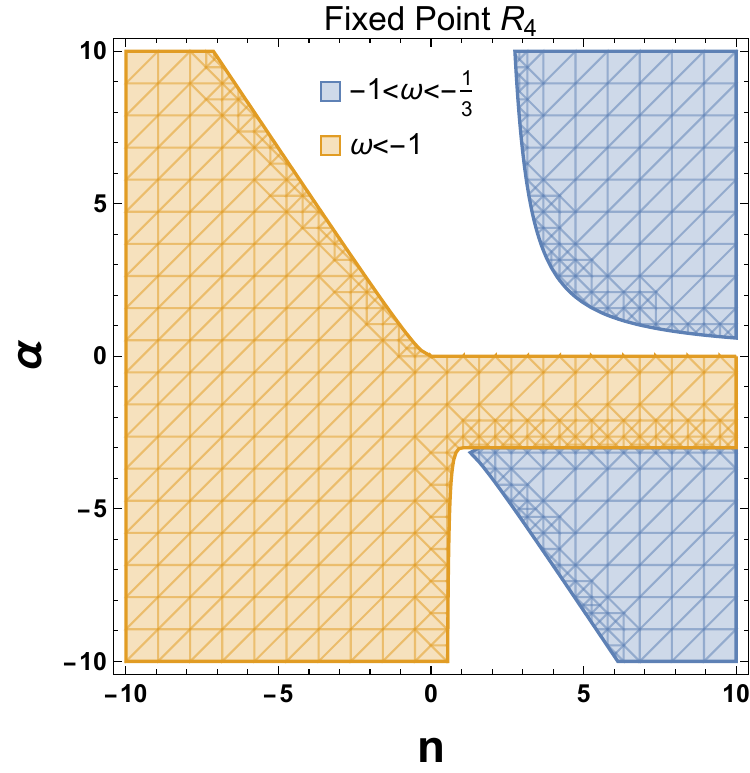}
         \caption{}
         \label{a}
     \end{subfigure}
     \hfill
     \begin{subfigure}[b]{0.45\textwidth}
         \centering
         \includegraphics[width=\textwidth]{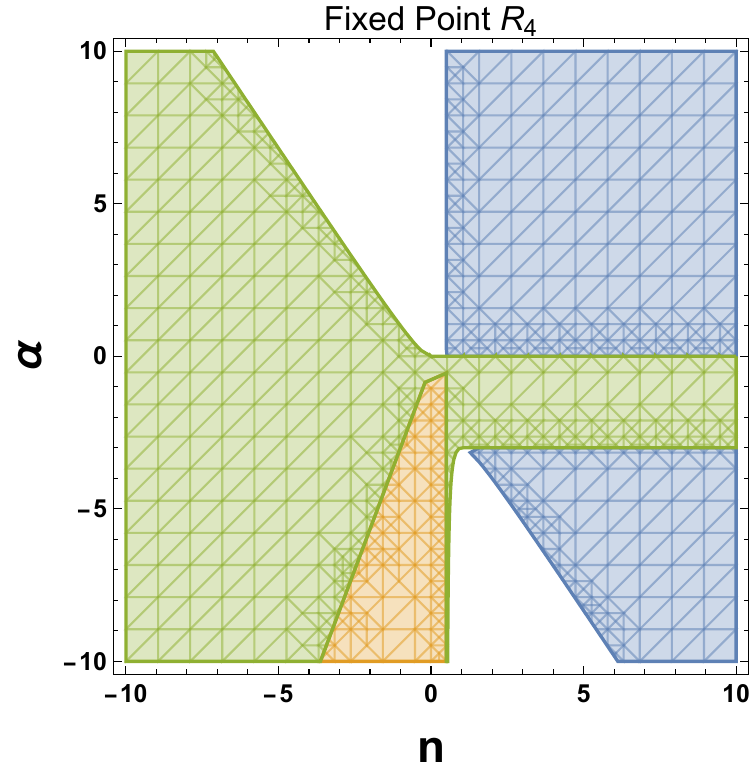}
         \caption{}
         \label{b}
     \end{subfigure}
     \caption{\justifying The region plot (a) shows the relationship between the model parameter $n$ and the interacting parameter $\alpha$ for the accelerating universe. The light blue region represents the quintessence-like behavior of an accelerating universe (i.e., $-1<\omega<-1/3$), while the golden region represents the phantom-like behavior of an accelerating universe (i.e., $\omega<-1$). The region plot (b) illustrates the stable, unstable, and saddle behaviors associated with the fixed point $R_4$.}
        \label{6}
\end{figure}

Corresponding to this fixed point, the matter density, DE density, and effective EoS parameters are obtained as 
 \begin{eqnarray*}
  &&  \Omega_m = \frac{-\alpha +\sqrt{-3 \alpha +9 n^2+6 \alpha  n}+(2 \alpha +3) n}{\alpha  (2 n-1)}\,, \\
  && \Omega_{DE} = \frac{-\sqrt{3} \sqrt{-\alpha +3 n^2+2 \alpha  n}-3 n}{2 \alpha  n-\alpha }\,,\,\,\text{and}\\
  && w = \frac{-3 \alpha -4 \alpha  n^2+2 \sqrt{-3 \alpha +9 n^2+6 \alpha  n}+(8 \alpha +6) n}{\alpha  (1-2 n)^2}\,.
\end{eqnarray*}
One can note that all three parameters depend on the model parameter $n$ as well as the interacting parameter $\alpha$. Check appendix \ref{Appendix} for calculations. The conditions for an accelerating universe and the stability are shown in Figs. \ref{a} and \ref{b}.

\subsubsection{$R_5$: Additional scaling solution point}\label{R5}
Corresponding to the fixed point $R_5$, that is
\begin{equation*}
    R_5: (x,y)=\left(\frac{\sqrt{3} \sqrt{-\alpha +3 n^2+2 \alpha  n}-3 n}{2 \alpha  n-\alpha },0\right)
\end{equation*}
the matter density, DE density, and effective EoS parameters are given by
 \begin{eqnarray*}
  &&  \Omega_m=\frac{3 n-\sqrt{-3 \alpha +9 n^2+6 \alpha  n}}{\alpha  (2 n-1)}+1\,, \\
  && \Omega_{DE} = \frac{\sqrt{-3 \alpha +9 n^2+6 \alpha  n}-3 n}{\alpha  (2 n-1)}\,,\,\,\\
  && w= \frac{-3 \alpha -4 \alpha  n^2-2 \sqrt{-3 \alpha +9 n^2+6 \alpha  n}+(8 \alpha +6) n}{\alpha  (1-2 n)^2}\,.
\end{eqnarray*}

\begin{figure}
     \centering
     \begin{subfigure}[b]{0.45\textwidth}
         \centering
         \includegraphics[width=\textwidth]{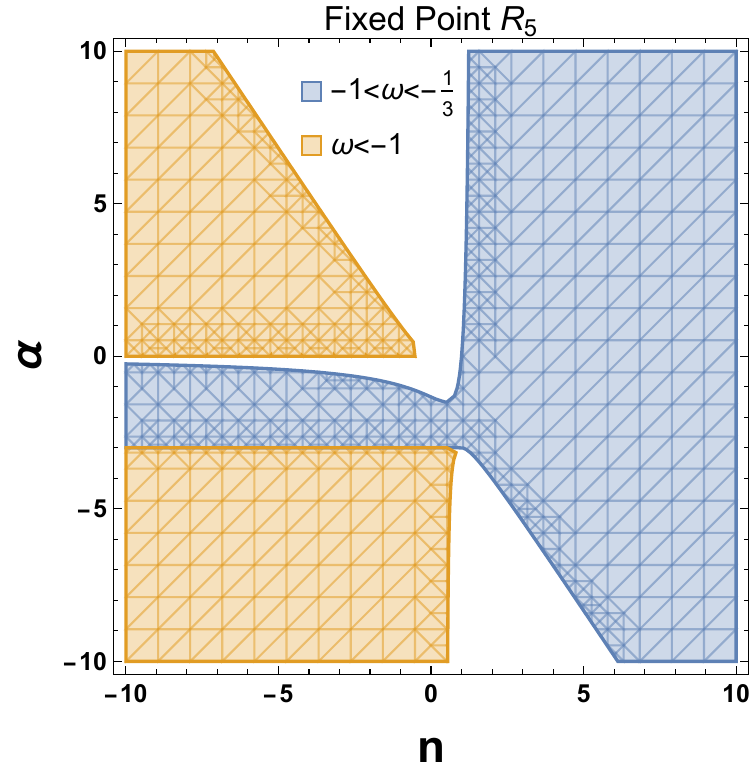}
         \caption{}
         \label{a1}
     \end{subfigure}
     \hfill
     \begin{subfigure}[b]{0.45\textwidth}
         \centering
         \includegraphics[width=\textwidth]{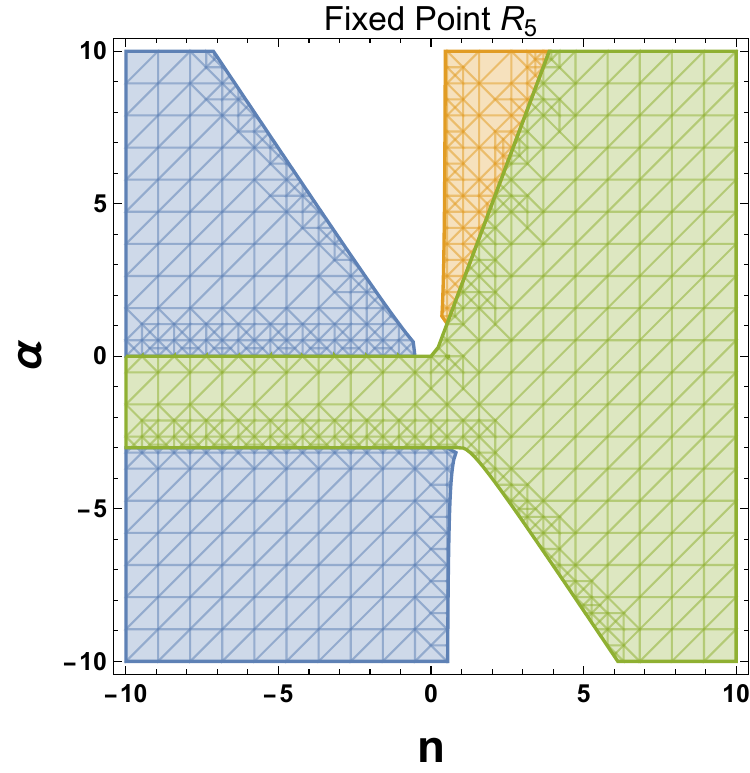}
         \caption{}
         \label{b1}
     \end{subfigure}
     \caption{\justifying The region plot (a) shows the relationship between the model parameter $n$ and the interacting parameter $\alpha$ for the accelerating universe. The light blue region represents the quintessence-like behavior of an accelerating universe (i.e., $-1<\omega<-1/3$), while the golden region represents the phantom-like behavior of an accelerating universe (i.e., $\omega<-1$). The region plot (b) illustrates the stable, unstable, and saddle behaviors associated with the fixed point $R_5$.}
        \label{8}
\end{figure}

The conditions for an accelerating universe and the stability are depicted in Figs. \ref{a1} and \ref{b1}. 

\begin{figure}[]
\centering
\includegraphics[scale=0.65]{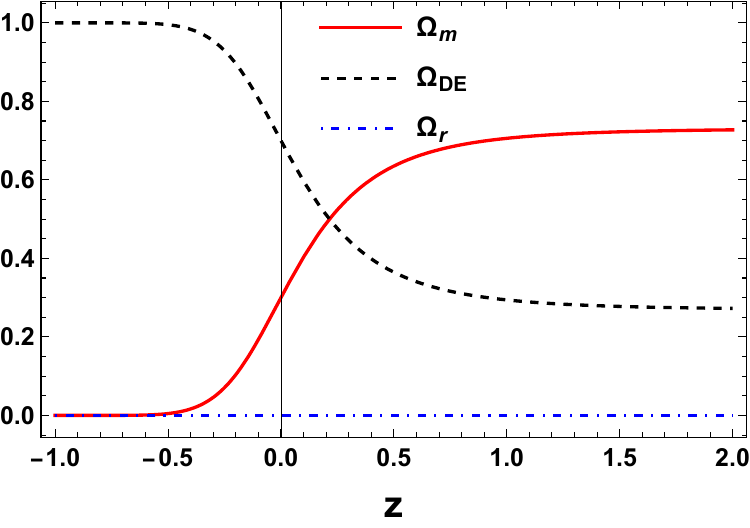} \hspace{0.2in}
\includegraphics[scale=0.65]{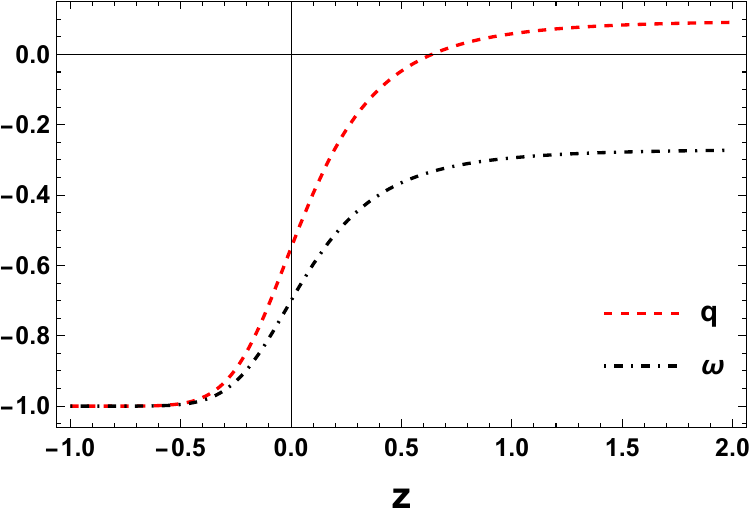}
\caption{\justifying The left panel displays density parameters for matter ($\Omega_m$), DE ($\Omega_{DE}$), and radiation ($\Omega_r$) as functions of redshift $z$. The right panel shows the deceleration parameter $q$ and the EoS parameter $\omega$ as functions of redshift $z$. These two panels illustrate the trajectory for the positive coupling parameter, which can transfer energy from DM to DE.}
\label{10}
\end{figure}

\begin{figure}[]
\centering
\includegraphics[scale=0.65]{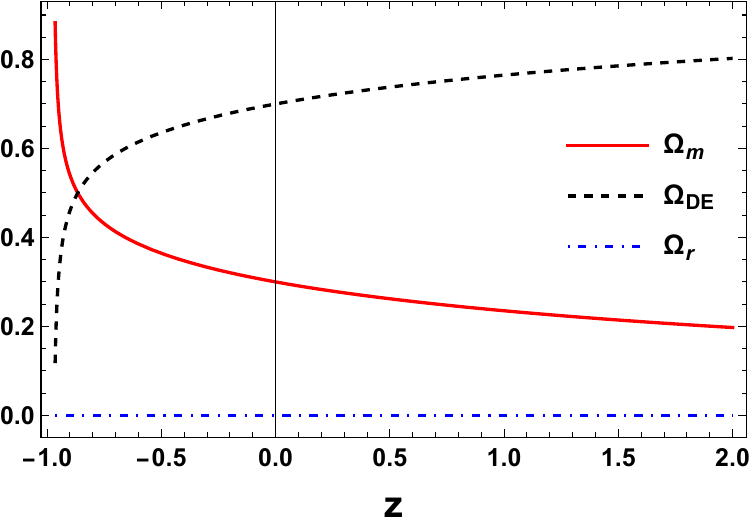} \hspace{0.2in}
\includegraphics[scale=0.65]{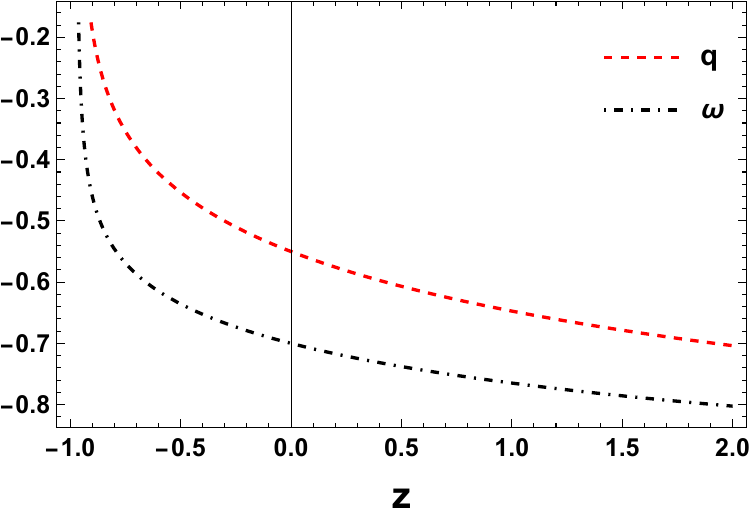}
\caption{\justifying The left panel displays density parameters for matter ($\Omega_m$), DE ($\Omega_{DE}$), and radiation ($\Omega_r$) as functions of redshift $z$. The right panel shows the deceleration parameter $q$ and the EoS parameter $\omega$ as functions of redshift $z$. These two panels illustrate the trajectory for the negative coupling parameter, which can transfer energy from DE to DM.}
\label{11}
\end{figure}

In Figures \ref{10} and \ref{11}, we illustrate the evolution of \(\Omega_m\), \(\Omega_r\), \(\Omega_{DE}\), \(q\), and \(\omega\) for different values of the model parameter \(n\) and the interaction parameter \(\alpha\). For \(n = 3/2\) and \(\alpha = 4\) in Figure \ref{10}, \(\alpha > 0\) indicates that the coupling term \(\mathcal{U} > 0\), signifying energy transfer from DM to DE. This figure shows that the universe was dominated by matter in the early stages and will be dominated by DE in later stages. Currently, the universe is dominated by DE, with parameters value \(\Omega_m = 0.3\), \(\Omega_r = 0.00005\), \(\Omega_{DE} = 0.7\), \(q_0 = -0.548\), and \(\omega_0 = -0.696\). For these values, the fixed points \(R_1\) and \(R_4\) are stable and represent the de Sitter and quintessence acceleration solutions, respectively.\\
In Figure \ref{11}, with \(n = 3/2\) and \(\alpha = -4\), \(\alpha < 0\) indicates that the coupling term \(\mathcal{U} < 0\), meaning energy transfers from DE to DM. This figure shows that the universe was dominated by DE in the early stages and will be dominated by DM in later stages. Currently, the universe remains dominated by DE, with parameters value \(\Omega_m = 0.3\), \(\Omega_r = 0.00005\), \(\Omega_{DE} = 0.7\), \(q_0 = -0.55\), and \(\omega_0 = -0.70\). For these values, the fixed points $R_4$ and $R_5$ have imaginary values, and the remaining point cannot exhibit stable behavior.

\begin{figure}[]
\centering
\includegraphics[scale=0.45]{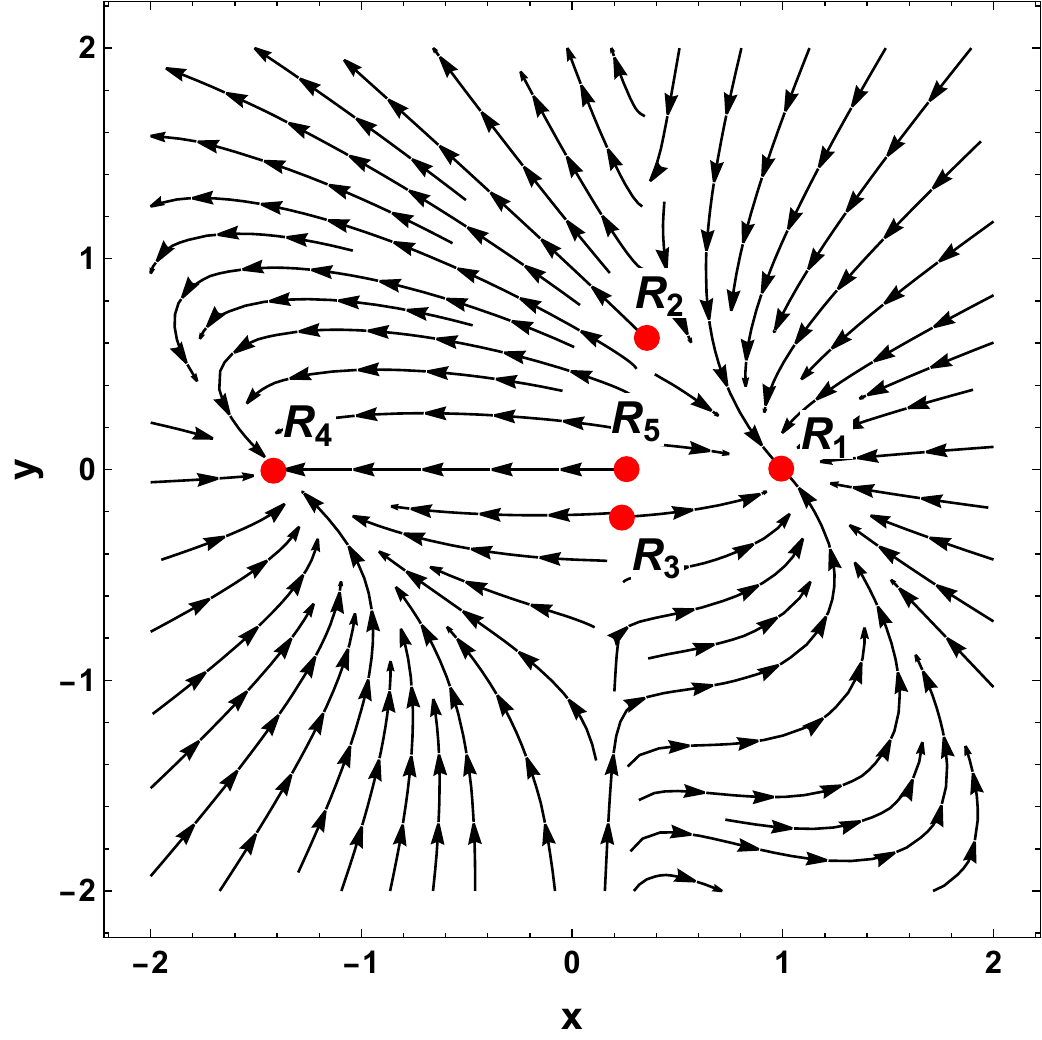} \hspace{0.2in}
\includegraphics[scale=0.45]{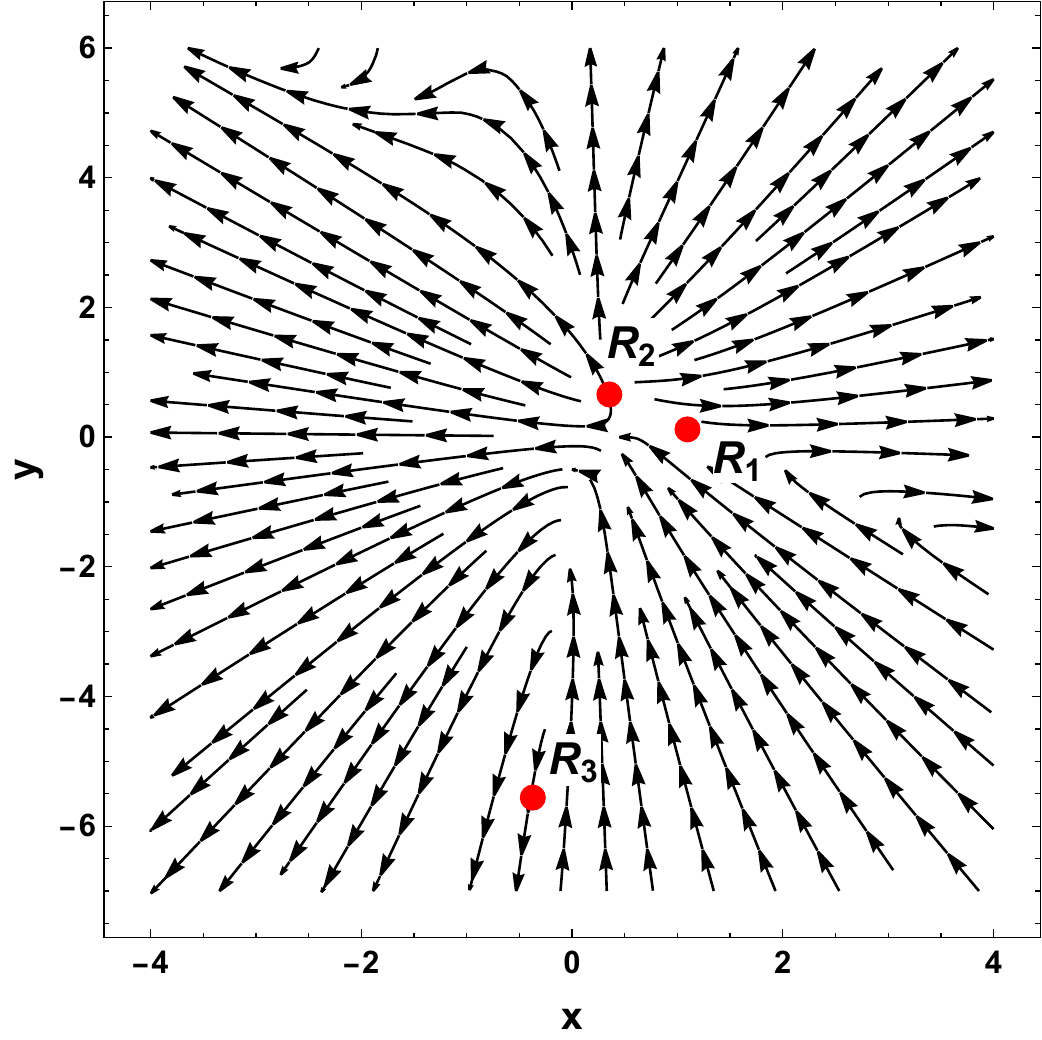}
\caption{Phase plot of $x$ versus $y$. These phase plots depict the stable behavior of $R_1$ and $R_4$ for $\alpha > 0$.}
\label{12}
\end{figure}

\section{Conclusion}
\label{section 5}
In this work, we have explored the behavior of interacting DE and DM within $f(Q)$ gravity, employing a standard framework of dynamical system analysis. We have considered the power-law $f(Q)=6\gamma\,H_0^2\left(\tfrac{Q}{Q_0}\right)^n$ model incorporating with two different forms of interacting DE and DM: $3\alpha H\rho_m$ and $\frac{\alpha}{3H}\rho_m \rho_{DE}$. The parameter $\alpha$ in the interacting terms plays a crucial role in determining the viable conditions and estimating the transition from the matter-dominated era (saddle point) to the DE-dominated era (stable node) in viable gravity models at late times. As a result, we have discovered fixed points that can be represented as the late-time accelerating universe in $f(Q)$ gravity. For the form of ${\cal U}=3\alpha H\rho_m$, we have illustrated the evolution of \(\Omega_m\), \(\Omega_r\), \(\Omega_{DE}\), \(q\), and \(\omega\) for different values of the model parameter \(n\) and the interaction parameter \(\alpha\). For \(n = 3/2\) and \(\alpha = 1/2\) and \(\alpha > 0\), we found that the coupling term \(\mathcal{U} > 0\), signifying energy transfer from DM to DE implying that the universe was dominated by matter in the early stages and would be dominated by DE in later stages. With the current data, fixed point \(P_1\) is stable and represents the de Sitter acceleration solution, while fixed point \(P_2\) is a saddle node, and \(P_3\) is an unstable node that cannot demonstrate universal acceleration. Moreover, we have considered another situation of which \(n = 3/2\) and \(\alpha = -2\), \(\alpha < 0\). We discovered for the coupling term \(\mathcal{U} < 0\) that energy transfers from DE to DM. In this case, the universe was dominated by DE in the early stages and will be dominated by DM in later stages. With the current data, fixed point \(P_3\) is stable and exhibits acceleration for the early and present universe but fails to show acceleration for late times, while fixed point \(P_1\) is a saddle node, and \(P_2\) is an unstable node.

For the form of ${\cal U}=\frac{\alpha}{3H}\rho_m \rho_{DE}$, the evolution of \(\Omega_m\), \(\Omega_r\), \(\Omega_{DE}\), \(q\), and \(\omega\) for different values of the model parameter \(n\) and the interaction parameter \(\alpha\) have been examined. We have considered \(n = 3/2\) and $\alpha=4$ (\(\alpha > 0\)) and found that the coupling term \(\mathcal{U} > 0\), signifying energy transfer from DM to DE. Our results show that the universe was dominated by matter in the early stages and would be dominated by DE in later stages. Using the observational data, the fixed points \(R_1\) and \(R_4\) were stable and would represent the de Sitter and quintessence acceleration solutions. Additionally, for \(n = 3/2\) and $\alpha=-4$ (\(\alpha < 0\)), we found that the coupling term \(\mathcal{U} < 0\), meaning energy transfers from DE to DM implying that the universe was dominated by DE in the early stages and would be dominated by DM in later stages. Employing the current data, the fixed points $R_4$ and $R_5$ have imaginary values, and the remaining point cannot exhibit stable behavior.

Although many viable $f(Q)$ models explain the DE problem in cosmology, our qualitative results from this work can serve as guidelines for more detailed studies. They can also be complementary constraints on viable $f(Q)$ models alongside other cosmological constraints on $f(Q)$ theories. Our framework, based on the cosmological dynamics of interacting DE and DM in $f(Q)$ gravity, constitutes the natural template beyond the standard gravity model, for example, Teleparallel gravity and Gauss-Bonnet gravity, or even more generalizations. Advancing our understanding requires developing new theoretical tools to analyze how interactions influence linear and non-linear regimes. Precisely determining an interaction kernel is crucial, as it could offer profound insights into fundamental physics, particularly the nature and properties of DM and DE.

\section*{\NoCaseChange{Data Availability Statement}}
There are no new data associated with this article.

\section*{\NoCaseChange{Acknowledgments}}
GNG acknowledges University Grants Commission (UGC), New Delhi, India for awarding a Senior Research Fellowship (UGC-Ref. No.: 201610122060). SA acknowledges the Japan Society for the Promotion of Science (JSPS) for providing a postdoctoral fellowship during 2024-2026 (JSPS ID No.: P24318). This work of SA is supported by the JSPS KAKENHI Grant (Number: 24KF0229). PKS acknowledges Anusandhan National Research Foundation, Department of Science and Technology, Government of India for financial support to carry out Research project No.: CRG/2022/001847 and IUCAA, Pune, India for providing support through the visiting Associateship program. We are very much grateful to the honorable referee and to the editor for the illuminating suggestions that have significantly improved our work in terms of research quality, and presentation. 
\appendix

\section{Appendix} \label{Appendix}
The conditions of stability for the fixed point $R_3$ are as follows:
\begin{itemize}
 \item For Stable Node:
\begin{eqnarray}
&& \frac{1}{8}<n\leq 0.4393\,\, \text{and}\,\, 0<\alpha <\frac{1}{3} (8 n-1).
\end{eqnarray}
\item For Unstable Node:
\begin{eqnarray}
&& n=0\,\, \text{and}\,\, \left(-\frac{1}{3}<\alpha <0\,\,\text{or}\,\, 0<\alpha \leq 1\right),\\
&& n<0\,\,\text{and}\,\,\left(\frac{1}{3} (8 n-1)<\alpha <2 n\right),\\
&& 0<n\leq \frac{1}{8}\,\,\text{and}\,\,\left(\frac{1}{3} (8 n-1)<\alpha <0\right),\\
&& n>\frac{1}{2}\,\,\text{and}\,\, 2 n<\alpha <\frac{1}{3} (8 n-1).
\end{eqnarray}
\item While the saddle point is otherwise.
\end{itemize}

The conditions of stability for the fixed point $R_4$ are as follows:
\begin{itemize}
\item For Stable node:
\begin{eqnarray}
&& 0<n\leq \frac{1}{8}\,\,\text{and}\,\,0<\alpha <-\frac{3 n^2}{2 n-1},\\ 
&& \frac{1}{8}<n<\frac{1}{5}\,\, \text{and}\,\, \frac{1}{3} (8 n-1)<\alpha <-\frac{3 n^2}{2 n-1},\\
&& \frac{1}{2}<n\leq 1\,\, \text{and}\,\, \alpha >0,\\
&& n>1\,\, \text{and}\,\, \left(-\frac{3 n^2}{2 n-1}<\alpha <-3\,\, \text{or}\,\, \alpha >0\right).
 \end{eqnarray}
 
\item For Unstable node:
\begin{eqnarray}
&& n\leq \frac{1}{8}\,\,\text{and}\,\,\alpha <\frac{1}{3} (8 n-1),\\
&& \frac{1}{8}<n<\frac{1}{2}\,\,\text{and}\,\, \alpha <0.
\end{eqnarray}

\item While the saddle point is otherwise.
\end{itemize}

The conditions of stability for the fixed point $R_5$ are as follows:

\begin{itemize}
    \item For Stable node:
    \begin{eqnarray}
    && n\leq 0\,\,\text{and}\,\, \left(\alpha <-3 \,\, \text{or}\,\, 0<\alpha <-\frac{3 n^2}{2 n-1}\right),\\
    && 0<n<\frac{1}{2}\,\,\text{and}\,\, \alpha <-3,\\
    && \frac{1}{2}<n<1\,\,\text{and}\,\, -\frac{3 n^2}{2 n-1}<\alpha <-3.
    \end{eqnarray}

    \item For Unstable node:
    \begin{eqnarray}
       && \frac{1}{5}<n<\frac{1}{2}\,\,\text{and}\,\, \frac{1}{3} (8 n-1)<\alpha <-\frac{3 n^2}{2 n-1},\\
       && n>\frac{1}{2}\,\,\text{and}\,\, \alpha >\frac{1}{3} (8 n-1).
    \end{eqnarray}
 \item While the saddle point is otherwise. 
\end{itemize}

\bibliography{ref}

\end{document}